# Verification of the generalized reduced-order particle-in-cell scheme in a radial-azimuthal E × B plasma configuration

F. Faraji [*][†], M. Reza[*], A. Knoll[*]

[*]Plasma Propulsion Laboratory, Department of Aeronautics, Imperial College London

[†]Corresponding Author (f.faraji20@imperial.ac.uk)

**Abstract**: In this article, we present an in-depth verification of the generalized electrostatic reduced-order particle-in-cell (PIC) scheme in a cross electric and magnetic field configuration representative of a radial-azimuthal section of a Hall thruster. The setup of the simulations follows a well-established benchmark case. The main purpose of this effort is to demonstrate that our novel PIC scheme can reliably resolve the complex two-dimensional dynamics and interactions of the plasma instabilities in the radial-azimuthal coordinates of a Hall thruster at a fraction of the computational cost compared to full-2D PIC codes. To this end, we first present the benchmarking of our newly developed full-2D PIC code. Next, we provide an overview of the reduced-order PIC scheme and the resulting "quasi-2D" code, specifying that the degree of order reduction in the quasi-2D PIC is defined in terms of the number of "regions" along the simulation's directions used to divide the computational domain. We compare the predictions of the quasi-2D simulation in various approximation degrees of the 2D problem against our full-2D simulation results. We show that, by increasing the number of regions in the Q2D simulations, the quasi-2D results converge to the 2D ones. Nonetheless, we also highlight that a quasi-2D simulation that provides a factor of 5 reduction in the computational cost resolves the underlying physical processes in an almost indistinguishable manner with respect to the full-2D simulation and incurs a maximum error of only about 10 % in the ion number density and about 5 % in the electron temperature.

## Section 1: Introduction

Nowadays, the E × B plasma discharges, where the applied magnetic field is perpendicular to the discharge current, have several important applications, from magnetrons for material processing to Hall thrusters for spacecraft plasma propulsion. The operation of the E × B technologies mainly relies on effectively impeding the cross-magnetic-field electron current. Nonetheless, the plasma in an E × B configuration is subject to strong anisotropies and gradients which lead to the excitation of various instabilities and turbulence, spanning over a broad range of spatial and temporal scales [1][2]. The instabilities and turbulent mechanisms are demonstrated through numerical [3][4] and experimental [5][6] efforts to induce notable cross-field transport of particles and energy. As a result, the global dynamics of the plasma discharge and, hence, the performance of the E × B devices are affected by these plasma processes. Therefore, acquiring an all-round knowledge of the instabilities' behaviour and their effects in E × B plasmas in conjunction with having predictive modelling tools available will greatly enhance the development cycle of the technologies reliant on these plasmas, enabling methodical and cost-effective design, optimization, and testing of the devices.

Although many insights into the excitation and evolution of the instabilities in E × B plasmas have been gathered over the past decades, a comprehensive knowledge basis of these phenomena and their interplays is still lacking [7]. The reason for this status-quo is two-fold: on one side, the dominant phenomena are three-dimensional and greatly multi-scale [7], are kinetic in nature [8], and their evolution and interactions are characterized by highly nonlinear processes [4][9]. On the other side, a comprehensive study of the complex dynamics of the instabilities and their associated effects on the plasma discharge requires a high-fidelity and computationally affordable 3D kinetic model capable of resolving the full-3D dynamics in real-size devices over real-world operational timeframes. Such a modelling tool is currently unachievable due to the significant computational burden of the existing kinetic simulation codes. Even lower dimension 2D kinetic simulations are prohibitively expensive for real-scale devices; for instance, a 2D PIC simulation of a Hall thruster for a representative period of 150 $\mu s$ simulated time, without any scaling of the physical or geometrical parameters, requires a run time of *about 1 year using 1500 CPUs* [10]. In view of the unavailability of a cost-effective 3D kinetic code, the main unresolved questions in the physics of E × B plasmas revolve around the 3D effects and interactions [7] as well as the mechanisms for energy transfer across instabilities spectrum over time that can lead to the formation of large-scale spatiotemporally coherent structures (i.e., inverse energy cascade and self-organization) [11]. Consequently, the puzzle of the dynamics of the particles and energy transport due to the instabilities and turbulence in various phases of their evolution has still important missing pieces.

The reduced-order, or quasi-multi-dimensional, PIC scheme is an innovative approach based on a dimensionality-reduction technique that reduces the computational cost of electrostatic kinetic simulations by 1 to 3 orders of magnitude [12][13]. As a result, it serves as a potential solution to the long-persisting challenge of the unavailability of a self-consistent, high-fidelity 3D kinetic code to reveal the full picture of the underlying plasma phenomena in E × B configurations. So far, the scheme is verified in detail in an axial-azimuthal Hall thruster configuration, demonstrating that it is able to approximate the full-2D results from conventional PIC codes with high accuracy and remarkable computational efficiency [13][14]. Moreover, the formulation underlying the reduced-order PIC scheme, which enables the decomposition of the multi-dimensional Poisson's equation into a system of coupled 1D ODEs, is proven to be mathematically consistent and applicable to solve general Poisson problems with arbitrary boundary conditions [13].

Based on the already observed promises in the axial-azimuthal configuration, and ahead of extending the reduced-order PIC to 3D, we evaluate, in this effort, the applicability and predictions' accuracy of the reduced-order PIC in another 2D E × B configuration, which is representative of a radial-azimuthal section of a Hall thruster. Prior research into the physics of plasma in the radial-azimuthal Hall thruster configuration has revealed the development of microscopic instabilities [15][16] and intricate interactions between the plasma sheath and the bulk plasma processes that affect the evolution of the excited kinetic instabilities [17]. Thus, verifying the "quasi-2D" PIC in the radial-azimuthal configuration allows us to ascertain further the abilities of the scheme in resolving closely coupled, highly two-dimensional phenomena, hence, gaining additional confidence that the generalized implementation of the reduced-order scheme serves as the required strong foundation upon which we can build to enable reliable and computationally efficient 3D simulations of the E × B plasmas.

Toward the above aim, we provide in Section 2 a brief review of the reduced-order quasi-2D PIC scheme and the corresponding kinetic code, IPPL-Q2D, developed at Imperial Plasma Propulsion Laboratory (IPPL). We also introduce our laboratory's recently developed full-2D conventional PIC code, IPPL-2D. Since the full-2D and the quasi-2D PIC codes share several similarities, the comparison of the quasi-2D results against our own 2D ones, rather than the reference benchmark results [18], allows us to unequivocally link the potential discrepancies to the reduced-order description of the problem inherent to the quasi-2D scheme. In Section 3, we provide an overview of the radial-azimuthal simulation setup and conditions. Moreover, we present the verification results of the IPPL-2D code against the benchmark in this section. We compare the predictions of the IPPL-Q2D code in various number-of-region approximations and those from the IPPL-2D in Section 4, discussing the results in terms of the radial distribution of the time-averaged plasma properties, ion number density and electron temperature, the time evolution of these plasma properties, the wavenumber content of the resolved azimuthal electric field fluctuations, and the 2D radial-azimuthal snapshots of several plasma parameters at two instances of the discharge evolution. In Section 5, we further evaluate the flexibility of the reduced-order PIC by investigating the effect that adopting non-uniform regions' extent in the quasi-2D simulations has on the results. Section 6 is dedicated to the analysis of the sensitivity of the quasi-2D results to the initial number of macroparticles per cell. Finally, in Section 7, we highlight the main outcomes of this study and outline the planned follow-up activities.

**Section 2: IPPL-2D and IPPL-Q2D particle-in-cell codes**

In this section, we first introduce the IPPL-2D code and then provide an overview of our reduced-order PIC, IPPL-Q2D. The difference between these two electrostatic particle-in-cell codes is limited to two aspects: first, the formulation underlying the Poisson solver and, second, the computations' grid that we use to deposit the particle-based information, such as the charge density, and from which we gather the grid-based information, like the electric field, onto particles' location. All other module implementations such as the random number generator, particles' push, and the diagnostics, are the same between the two codes.

**2.1. Introducing the IPPL-2D particle-in-cell code**

IPPL-2D is an explicit, conventional fully kinetic particle-in-cell code whose overall structure and function implementations follow the standard particle-in-cell method, as explained in Refs. [19][20].

The code is written in Julia language [21] and uses the built-in Random Number Generator (RNG) function, which follows Xoshiro256++ algorithm [22] by default. The sampling of the distribution function for macroparticles' load is done using the Box-Muller algorithm [23]. Linear weighting is used to deposit particle-based information onto the 2D grid nodes and to gather the grid-based information onto particles' position. Particle's push function is based on the classic leap-frog scheme [19] with the position of the magnetized electrons being updated following the Boris method [24]. The electric potential is resolved by numerically solving the 2D Poisson's equation using Julia's built-in direct matrix solve algorithm based on the LU decomposition.

**2.2. Overview of the generalized reduced-order PIC scheme and the IPPL-Q2D PIC code**

As mentioned in Section 1, the reduced-order PIC scheme, described in detail in Refs. [13] and [25], is a novel kinetic plasma modeling approach to significantly lower the computational resource requirement of the traditional, multi-dimensional PIC codes. In this PIC scheme, the computational domain is decomposed into multiple rectangular (in 2D) or cubic (in 3D) "regions", which can be thought of as the discretization of the domain using a "coarse" grid (Figure 1, left). Within any specific region, denoted as $\Omega$ on the left-hand side of Figure 1, we discretize each simulation dimension separately using 1D "elongated" cells (orange-shaded elements in Figure 1, right). The criterion for the cell size along the x and y direction, $\Delta x$ and $\Delta y$, is the same as that in a conventional 1D PIC, i.e., smaller than the Debye length. This fine-discretization step yields a decoupling between the different coordinates in each region, allowing us to reduce the number of cells and, hence, the required total number of macroparticles from $O(N^2)$ to $O(2N)$ in a "quasi-2D" (or from $O(N^3)$ to $O(3N)$ in a "quasi-3D") simulation. The vertical and horizontal cells in each region (Figure 1, right) capture separately the variations of the plasma properties along the x and y directions, respectively.

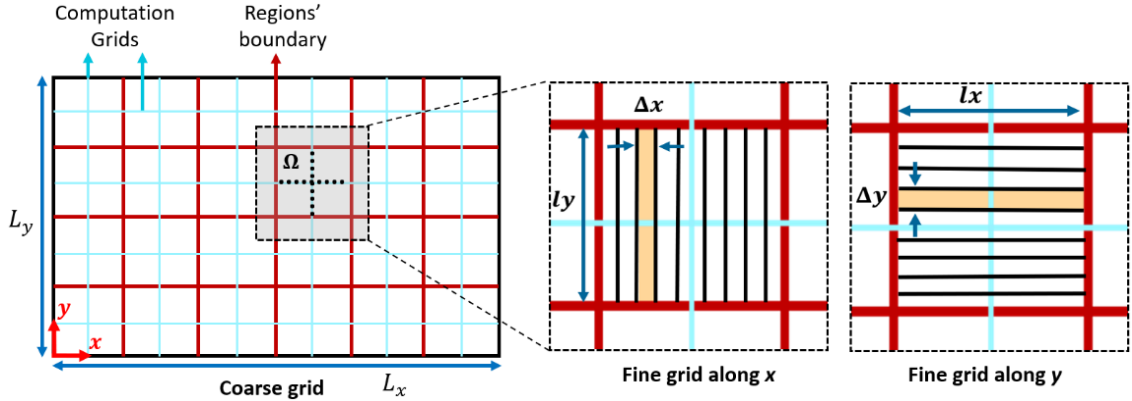

Figure 1: Schematic of the domain decomposition corresponding to the reduced-order PIC scheme; (left) decomposition into multiple "regions" using a coarse grid, (right) the 1D computational cells for discretization of each region along the x and y directions.

The dimensionality-reduction formulation underpinning the generalized reduced-order PIC scheme [13] is formulated in a flexible way that allows the user to determine the desired level of predictions' accuracy and, hence, to define the order of approximation. This, in turn, changes the required computational resources [13]. In other words, the accuracy and the computational cost of a reduced-order simulation depend on the fineness of the coarse grid (Figure 1, left), i.e., the number of regions to be used. Therefore, our quasi-multi-dimensional PIC scheme offers a unique flexibility to have a trade-off between fidelity and computational gain of the simulation depending on its specific purpose.

The IPPL-Q2D code, which is based on the reduced-order PIC scheme, features an electric potential field solver, which we have termed "Reduced-Dimension" Poisson Solver, or RDPS [13]. The RDPS numerically solves a system of 1D Poisson's equations along the simulation coordinates obtained from our dimensionality-reduction formulation, Eqs. 1 and 2, using the same Julia's built-in direct matrix solve algorithm as that of the IPPL-2D. The detailed derivation of Eqs. 1 and 2 as well as the verification results of the RDPS for generic 2D Poisson problems are reported in Ref. [13].

In Eqs. 1 and 2, $\phi^x$ and $\phi^y$ are 1D potential functions along the x and y directions, respectively, that are used to approximate the 2D potential $\phi(x,y)$ in each region $\Omega$. $l_x$ and $l_y$ are the horizontal and vertical extents of each region, $x_i$ is the x-coordinate of each computational node $i$ along a horizontal computation grid, and $y_j$ is similarly the y-coordinate of each node $j$ along a vertical computation grid. In addition, $x_g$ and $y_g$ are the locations of the y and x computation grid, respectively, with respect to the origin, $\epsilon_0$ is the permittivity of free space, and $\rho(x,y)$ is the 2D charge density distribution in region $\Omega$.

$$\left(\frac{d^2\phi^x}{dx^2}\Big|_{x=x_i}\right)l_y + \frac{\partial\phi^x}{\partial y}\Big|_{y=y_g+\frac{l_y}{2}} - \frac{\partial\phi^x}{\partial y}\Big|_{y=y_g-\frac{l_y}{2}}$$

$$= -\frac{1}{\epsilon_0}\int_{y_g-\frac{l_y}{2}}^{y_g+\frac{l_y}{2}}\rho(x,y)dy - \frac{\partial\phi^y}{\partial y}\Big|_{y=y_g+\frac{l_y}{2}} + \frac{\partial\phi^y}{\partial y}\Big|_{y=y_g-\frac{l_y}{2}}$$

$$-\left(\int_{y_g-\frac{l_y}{2}}^{y_g+\frac{l_y}{2}}\left(\frac{\partial\phi^y}{\partial x}\Big|_{x=x_i+\frac{\Delta x}{2}}\right)dy - \int_{y_g-\frac{l_y}{2}}^{y_g+\frac{l_y}{2}}\left(\frac{\partial\phi^y}{\partial x}\Big|_{x=x_i-\frac{\Delta x}{2}}\right)dy\right)\frac{1}{\Delta x},$$

(Eq. 1)

$$\left(\frac{d^2\phi^y}{dy^2}\Big|_{y=y_j}\right)l_x + \frac{\partial\phi^y}{\partial x}\Big|_{x=x_g+\frac{l_x}{2}} - \frac{\partial\phi^y}{\partial x}\Big|_{x=x_g-\frac{l_x}{2}}$$

$$= -\frac{1}{\epsilon_0}\int_{x_g-\frac{l_x}{2}}^{x_g+\frac{l_x}{2}}\rho(x,y)dx - \frac{\partial\phi^x}{\partial x}\Big|_{x=x_g+\frac{l_x}{2}} + \frac{\partial\phi^x}{\partial x}\Big|_{x=x_g-\frac{l_x}{2}}$$

$$-\left(\int_{x_g-\frac{l_x}{2}}^{x_g+\frac{l_x}{2}}\left(\frac{\partial\phi^x}{\partial y}\Big|_{y=y_j+\frac{\Delta y}{2}}\right)dx - \int_{x_g-\frac{l_x}{2}}^{x_g+\frac{l_x}{2}}\left(\frac{\partial\phi^x}{\partial y}\Big|_{y=y_j-\frac{\Delta y}{2}}\right)dx\right)\frac{1}{\Delta y}.$$

(Eq. 2)

It is worth pointing out that, at each simulation time step, the derivative terms on the right-hand side of Eqs. 1 and 2 are obtained from the solutions of $\phi^x$ and $\phi^y$ from the previous time step [13].

The RDPS is one of the differences between IPPL-Q2D and IPPL-2D code. It provides the solution of the electric potential and, hence, the electric field, on the reduced-dimension computation grids illustrated as the blue lines in Figure 1. As a result, in IPPL-Q2D code, these are the grids that we use for the deposition of the particles' charge density and for gathering the electric field components to advance the particles' position. This is the second and last difference between our quasi-2D and full-2D codes.

### Section 3: Description of the radial-azimuthal simulation setup and IPPL-2D benchmarking

The setup of the full-2D and quasi-2D simulations presented in this work exactly follows that of the well-established radial-azimuthal benchmark case by Villafana et al. [18]. We provide an overview of the simulation setup and conditions in Section 3.1. Next, in Section 3.2, we present and compare the results from our full-2D code against the reference benchmark results.

#### 3.1. Simulation domain and conditions

Figure 2 illustrates the domain for the 2D and quasi-2D simulations, which is a Cartesian $(x-z)$ plane with $x$ along the radial direction and $z$ along the azimuthal coordinate. The $y$-direction represents the axial coordinate. A constant radial magnetic field $(B_x)$ with the intensity of $20\ mT$ and a constant axial electric field $(E_y)$ with the magnitude of $1\times10^4\ Vm^{-1}$ is applied. A summary of the initial plasma properties and the computational parameters used for the simulations is provided in Table 1.

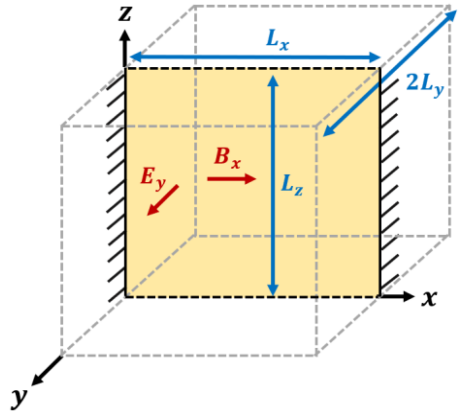

Figure 2: Schematics of simulation's coordinate system, the computational plane, and the applied constant electric and magnetic fields

| Parameter | Value [unit] |
|---|---|
| **Computational parameters** | |
| Domain length ($L_x = L_z$) | 1.28 [cm] |
| Virtual axial length ($L_y$) | 1 [cm] |
| Cell size ($\Delta x = \Delta z$) | 50 [μm] |
| Number of cells in each direction ($N_i = N_j$) | 256 |
| Time step ($ts$) | $1.5 \times 10^{-11}$ [s] |
| Total simulated duration ($t_{sim}$) | 30 [μs] |
| Initial number of macroparticles per cell ($N_{ppc}$) | 100 |
| **Physical parameters** | |
| Initial plasma density ($n_{i,0}$) | $5 \times 10^{16}$ [m$^{-3}$] |
| Initial electron temperature ($T_{e,0}$) | 10 [eV] |
| Initial ion temperature ($T_{i,0}$) | 0.5 [eV] |
| Axial electric field ($E_y$) | 10,000 [Vm$^{-1}$] |
| Radial magnetic field ($B_x$) | 0.02 [T] |
| Potential at walls ($\phi_w$) | 0 [V] |

Table 1: Summary of the computational and physical parameters used for the radial-azimuthal 2D and quasi-2D simulations

Initially, the electrons and ions are sampled from a Maxwellian distribution at 10 eV and 0.5 eV, respectively, and are loaded uniformly over the $(x - z)$ plane at exactly the same positions. In order to mimic the downstream convection of the azimuthal waves and to limit the growth of the particles' energy in the simulations [26], the approach of Ref. [18] is pursued, and a virtual axial extent with the length of $L_y = 1$ cm is considered. The particles crossing a domain's boundary along the axial direction are resampled from a Maxwellian at their initial load temperature, i.e., 10 eV for the electrons and 0.5 eV for the ions. These particles are re-injected on the simulation plane, maintaining their azimuthal and radial positions.

Referring to Figure 2, the two radial extremes of the domain, which represent the channel walls of a Hall thruster, are grounded, and the Secondary Electron Emission from the walls is neglected. As a result, a Dirichlet Boundary condition is applied for the electric potential at $x = 0$ and $x = L_x$, and any particle hitting a wall is removed from the simulation. The domain along the azimuthal direction is periodic. Hence, a periodic potential boundary condition is implemented, and the particles exiting the azimuthal extent at one end of the domain are re-introduced into the domain from the opposite end.

The simulation case is completely collisionless. However, to compensate for the flux of particles lost to the walls, following the benchmark's approach [18], a cosine-shaped ionization source, covering 86 % of the radial extent of the domain, from $x = 0.09$ cm to $x = 1.19$ cm, is imposed. The source is uniform along the azimuthal direction. The peak of the ionization source is $S_0 = 8.9 \times 10^{22}\ m^{-3}s^{-1}$, which corresponds to an axial current density of 100 $Am^{-2}$. The electron-ion pairs injected at each time step due to the ionization source are sampled from a Maxwellian at the respective initial temperature of each species.

### 3.2. Verification of the IPPL-2D against the radial-azimuthal benchmark results

Figure 3 shows the comparison between the radial distributions of the time-averaged plasma properties, ion number density and electron temperature, from IPPL-2D and from the benchmark's 2D code, CERFACS [18]. Referring to Ref. [18], the participating codes in the radial-azimuthal benchmarking activity had a maximum prediction error of $\pm$ 2.5 % in ion number density and electron temperature with respect to the CERFACS code.

Thus, we also took the results from CERFACS as the reference to verify our 2D results. It is seen in Figure 3 that the time-averaged radial distributions from IPPL-2D are almost identical to those from CERFACS, with only a negligible difference of about 0.5 % observable in the zoomed-in view on the ion number density profile (Figure 3(a)).

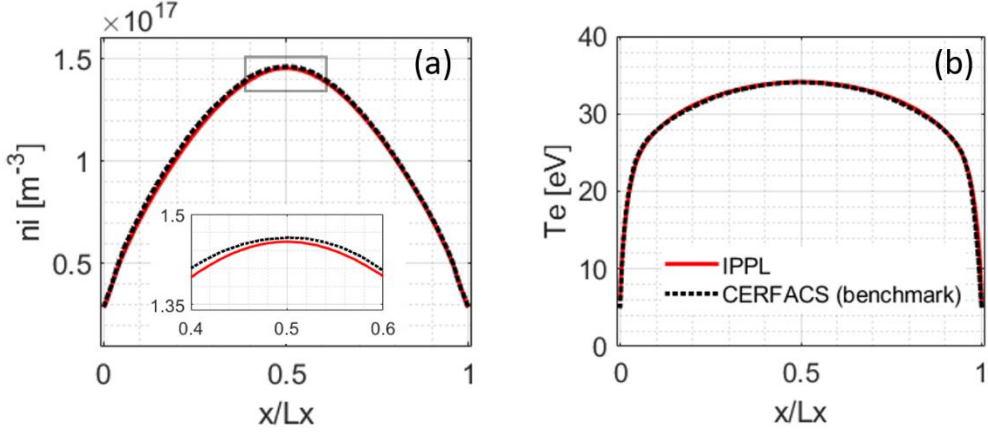

Figure 3: Comparison between the time-averaged radial profiles of (a) ion number density ($n_i$) and (b) electron temperature ($T_e$), obtained from the IPPL-2D and the CERFACS code [18]. The profiles are averaged over 25-30 $\mu s$.

The time evolution of the spatially averaged ion number density and the radial component of the electron temperature from IPPL-2D is shown in Figure 4. The radial electron temperature ($T_{ex}$) is obtained from Eq. 3, whereas the azimuthal electron temperature ($T_{ez}$), to be used for the discussions in Section 5, is given by Eq. 4. In Eqs. 3 and 4, $m_e$ is the electron mass, $e$ is the unit charge, $n$ is the number density, and $V_{de,x}$ and $V_{de,z}$ are, respectively, the electrons' radial and azimuthal drift velocity. $f_e(\boldsymbol{v})$ is the electrons' velocity distribution function.

$$T_{e,x} = \frac{m_e}{en}\left(\int_{-\infty}^{+\infty} v_x^2 f_e(\boldsymbol{v}) dv_x - nV_{de,x}\right), \qquad \text{(Eq. 3)}$$

$$T_{e,z} = \frac{m_e}{en}\left(\int_{-\infty}^{+\infty} v_z^2 f_e(\boldsymbol{v}) dv_z - nV_{de,z}\right). \qquad \text{(Eq. 4)}$$

Comparing the evolution plots of the plasma properties in Figure 4 with those published in Fig. 7 of the benchmarking article [18], we found that the resolved trends from IPPL-2D are very similar to those from the 2D codes that participated in the benchmark, both over the first 3 $\mu s$ of the simulation (Figure 4(a)) and over the entire simulation duration (Figure 4(b)).

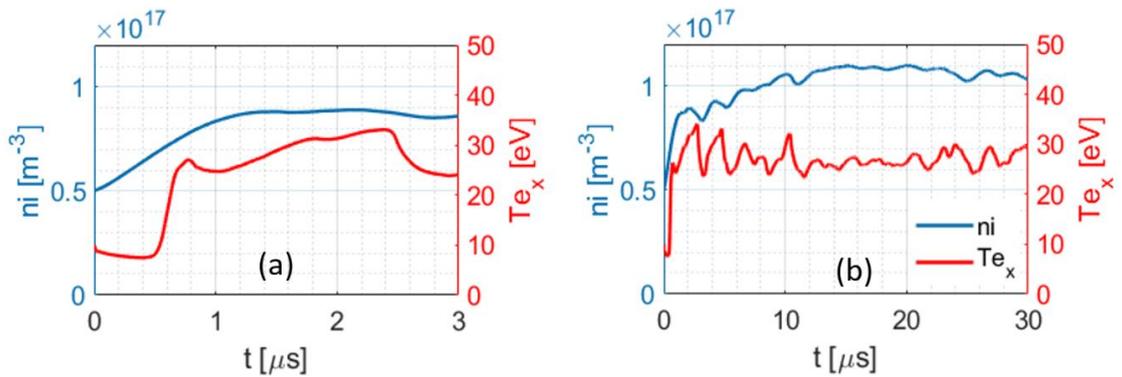

Figure 4: Time evolution of the ion number density ($n_i$) and the radial component of the electron temperature ($T_{e,x}$) over (a) the first 3 $\mu s$ of the simulation time and, (b) over the entire simulation duration. The left y-axis in both plots corresponds to the ion number density, whereas the right y-axis represents the radial electron temperature.

According to the above results and discussions, we verified that our full-2D PIC code provides predictions in the radial-azimuthal configuration that are highly consistent with the other benchmarked conventional full-2D codes in the E × B plasma community. Consequently, the comparison of the quasi-2D simulations' results against our full-2D ones, as presented in the next section, is justified.

# Section 4: Comparison between IPPL-Q2D predictions in various number-of-region approximations and IPPL-2D results

We performed a series of quasi-2D radial-azimuthal simulations using different orders of approximation of the 2D problem, which, as mentioned in Section 2.2, amounts to using various number of regions. In all of the quasi-2D simulations discussed in this section, an equal number of regions is used along the radial and azimuthal directions. For reference, we also report, in this section, the results from a 1D radial and a 1D azimuthal PIC simulation that we carried out using the same conditions and setup of the 2D and quasi-2D simulations (Section 3.1) but adapted accordingly for the 1D simulations. In the 1D radial case, the same ionization source as that described in Section 3.1 was used. However, for the 1D azimuthal simulation, noting that the plasma density remains constant throughout the simulation, the initial plasma density ($n_{i,0}$) was increased to $1 \times 10^{17}\ m^{-3}$, which approximately corresponds to the radially averaged density value from the 2D simulation (Section 3.2).

Figure 5 presents the time-averaged radial distributions of the ion number density ($n_i$) and the electron temperature ($T_e$) from the 1D radial simulation, from the quasi-2D simulations at "low" and "high" approximation orders, and from our 2D simulation. The "low-order" quasi-2D simulations correspond to the number of regions of 1, 3, and 5, whereas, for the "high-order" simulations, we used 10, 25, 50, and 100 regions.

Looking at the radial profiles of the plasma properties from the high-order quasi-2D simulations (Figure 5(c) and (f)), it is evident that, by increasing the order of approximation, the quasi-2D results converge to the full-2D ones at the 100 number-of regions limit. Indeed, the 100-region quasi-2D simulation results show an error of about 2.5% in peak ion number density and in peak electron temperature with respect to the 2D results. This is deemed a reasonably low error considering that it is consistent with the error observed among various 2D codes that participated in the radial-azimuthal benchmark activity [18].

Referring now to the plots (b) and (e) in Figure 5, it is observed that, by increasing the number of regions from 1 to 5, the ion number density profile becomes consistently more similar to the 2D one, whereas the electron temperature profile first shows an increase in values and a deviation in distribution compared to the 2D results from 1 region to 3 regions before starting to approach the 2D profile at 5 regions.

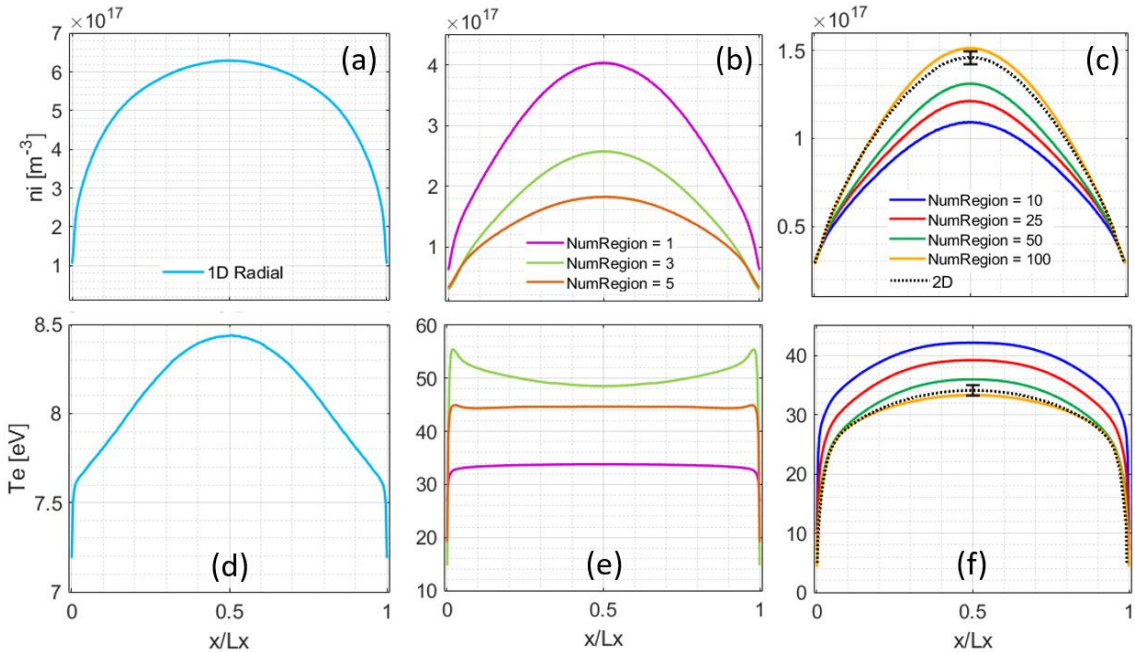

Figure 5: Radial profiles of the ion number density (first row) and the electron temperature (second row) averaged over 25-30 $\mu s$. Plots (a) and (d) are from the 1D radial simulation, (b) and (e) from low-order quasi-2D simulations, and (c) and (f) from high-order quasi-2D simulations. The reference 2D profiles are shown in plots (c) and (f) with a $\pm\ 2.5\ \%$ error band superimposed at the peak values. Note the difference in the y-axis scale among plots (a) to (c) and (d) to (f); the scales are chosen in this manner to allow distinguishing more clearly the time-averaged profiles from various simulations.

An interesting point to highlight is that the "single-region" quasi-2D simulation, which resolves an average effect of the radial and azimuthal processes [14], predicts a value of the electron temperature at the mid radial location ($x/L_x = 0.5$) that is very close to the corresponding value from the 2D simulation. However, the radial $T_e$ distribution from the single-region simulation is rather different from the 2D one. In addition, the predicted ion number density profile is also quite different in terms of values compared to the 2D result. Regardless of these discrepancies, comparing the single-region profiles against those from the 1D radial simulation (Figure 5(a) and (d)), it is seen that the single-region simulation provides a good improvement to the 1D radial results. In particular, the average mutual effects of the physical mechanisms along the azimuthal and radial directions are captured by the single-region simulation. This can be noticed by comparing the 1D radial and the single-region profiles where an increase in the electron temperature due to particles' heating by the azimuthal waves and a corresponding decrease in the ion number density is observed.

Figure 6 shows the time evolution of the ion number density and the radial electron temperature, averaged over the entire domain, from the 1D radial, quasi-2D, and the full-2D simulations. First, comparing the 1D radial and the single-region quasi-2D simulation, we notice that resolving an average mutual effect of the azimuthal and radial processes by the single-region simulation also affects the temporal variations of the plasma properties. In fact, in the 1D radial simulation, the ion number density (Figure 6(a)) approaches an asymptotic value near the end of the simulation duration, and the radial electron temperature (Figure 6(d)) drops to a value of about 3 eV very early into the simulation. In contrast, in the single-region simulation, the ion number density (Figure 6(b)) becomes asymptotic within the last 10 $\mu s$ of the simulation at a value 0.6 times that in the 1D radial case. In addition, the radial electron temperature shows a relatively small decrease from the initial load temperature of 10 eV and then oscillates around a mean value of 7.5 eV.

The reason for the above observations is the specific coupling that exists between the electron temperature and the number density in the adopted radial-azimuthal simulation case. In this regard, when $T_{ex}$ increases, an increased flux of particles is lost to the walls. Now, since the simulation setup features a constant ionization source that does not receive feedback from the system's energy and injects electron-ion pairs at a constant rate, the plasma density will consequently decrease. Accordingly, in the 1D radial case, after the initial transient where the tail of the electrons' radial energy distribution function (EDF) is depleted due to being lost to the wall and, hence, the radial electron temperature decreases, the absence of any mechanism to thermalize the electron population and, thus, to replenish the EDF tail, implies that the flux of particles with enough energy to overcome the sheath potential drop decreases. As a result, the ion number density evolution shows the increasing behavior seen in plot (a) of Figure 6. In the single-region simulation, the excitation of the azimuthal waves limits the decrease in the radial electron temperature which, as explained above, limits the increase in the ion number density.

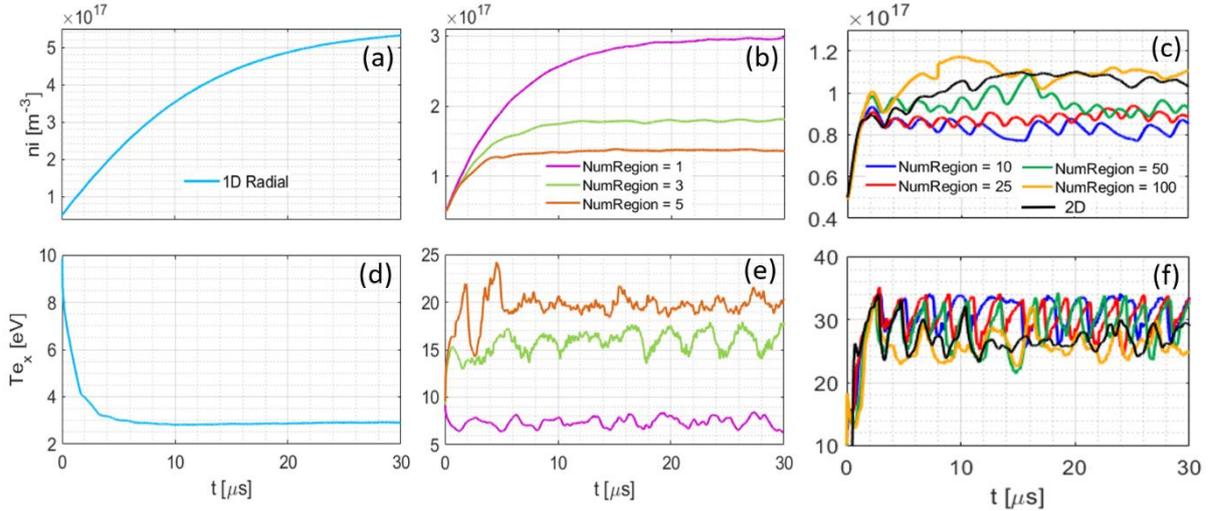

Figure 6: Time evolution of the ion number density (first row) and the radial electron temperature (second row) obtained from the performed simulations; (a) & (d) 1D-radial, (b) & (e) quasi-2D in low number-of-regions, and (c) & (f) in high number-of-regions approximation limit. The 2D simulation results are shown in the plots (c) & (f). Note the difference in the y-axis scale among plots (a) to (c) and (d) to (f); the scales are chosen in this manner to allow distinguishing more clearly the time evolutions from various simulations.

As the second point concerning the plots in Figure 6, the high-order quasi-2D simulations (Figure 6(c) and (f)) have resolved the time evolution of the radial electron temperature in a consistent manner compared to the 2D simulation. The temporal variations in $T_{ex}$ from all quasi-2D simulations feature a characteristic oscillation similar to that visible in the 2D result, which is associated with the periodic growth and damping of the Modified Two-Stream Instability (MTSI) [18]. Due to the coupling between the plasma density and the electron temperature in the present simulation case, which was explained in the preceding paragraph, the ion number density trends also show an oscillatory behavior over time (Figure 6(c)). In this regard, as the number of regions is increased from 10 to 100, the ion density time evolutions become increasingly more similar to that of the 2D, with 50- and 100-region simulations showing the highest degree of similarity overall.

Looking now at the plots (b) and (e) in Figure 6, we notice that as the approximation order in the low number-of-region quasi-2D simulations is increased from 1 to 5, the time to reach steady state for the ion number density decreases. Moreover, the mean value of the radial electron temperature oscillations becomes more comparable with the high-order quasi-2D and the full-2D ones from the single-region to 5-region simulation.

In Figure 7, we present the time evolution of the ion number density and the radial electron temperature from the high-order quasi-2D and the full-2D simulations over the first 3 $\mu s$ of the simulation. It is observed that, whereas the quasi-2D simulations have all resolved the initial transient in the ion number density rather consistently with the 2D simulation, the evolution trend of the radial electron temperature is quite different from the 2D one over approximately the first 1.5 $\mu s$ of the simulation. This difference highlights that the reduced-order simulations did not properly resolve the very initial evolution of the discharge, most likely due to the highly small-scale physical interactions occurring over the first few microseconds of the simulation as explained in Ref. [18]. Nevertheless, noting that the quasi-2D results are consistent with the 2D ones from about 2 $\mu s$ onward, we believe that the observed discrepancy here does not undermine the applicability and/or reliability of the reduced-order scheme, which is primarily intended to enable self-consistent kinetic simulations over long durations, which is currently not feasible using conventional full-2D codes without any simulation speed-up technique [27].

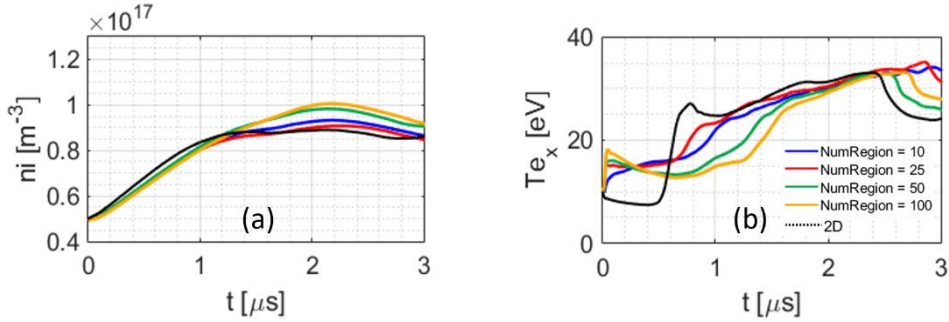

Figure 7: Time evolution of (a) ion number density and (b) radial electron temperature from the high-order quasi-2D and the 2D simulations over the first 3 $\mu s$ of the simulation.

The 1D Fast Fourier Transforms (FFT) of the azimuthal electric field signal, averaged over three time intervals along the simulation duration, from the 1D azimuthal simulation, the quasi-2D simulations in low- and high-order approximation limits, and from the 2D simulation are shown in Figure 8. The approach pursued to obtain these spatiotemporally averaged FFT plots is similar to that explained in detail in Refs. [15][18].

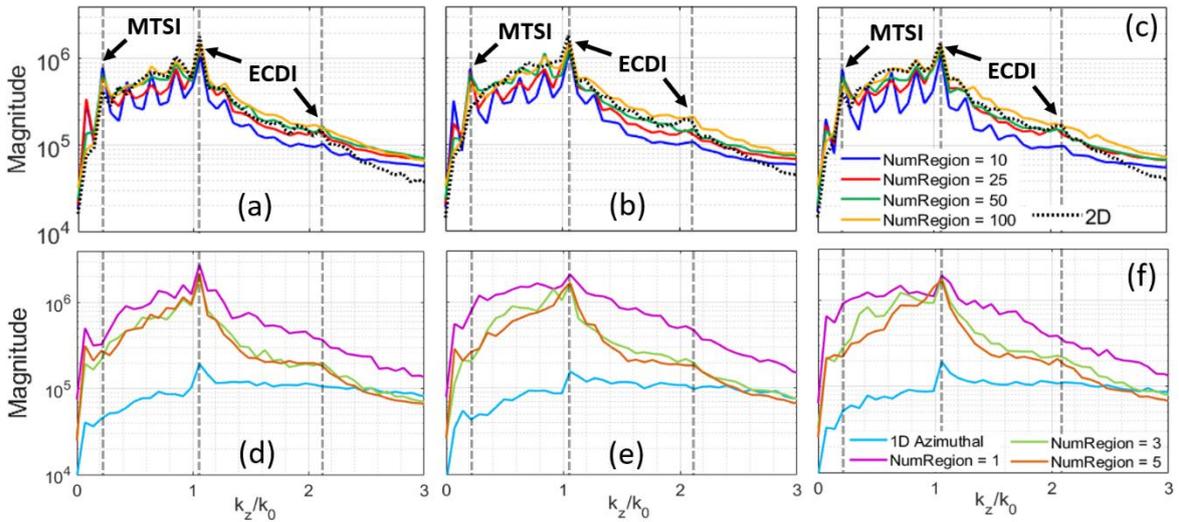

Figure 8: 1D FFTs of the azimuthal electric field signal averaged over all radial positions and over three time intervals; (a) & (d) 5-10 $\mu s$, (b) & (e) 15-20 $\mu s$, and (c) & (f) 25-30 $\mu s$. The top row is related to the quasi-2D simulations with high number-of-regions and the 2D simulation; the bottom row is related to the quasi-2D simulations with low number-of-regions and the 1D azimuthal simulation.

The x-axis in the plots of Figure 8 represents the normalized azimuthal wavenumber ($k_z/k_0$), with the normalization factor ($k_0$) being the fundamental resonance wavenumber of the Electron Cyclotron Drift Instability (ECDI) defined as $k_0 = \frac{\Omega_{ce}}{v_{de}}$ [18], where $\Omega_{ce}$ is the electron cyclotron frequency, and $v_{de}$ is the electrons' azimuthal drift velocity.

Referring to the top-row plots in Figure 8, which illustrate the FFTs of the azimuthal electric field from the high-order quasi-2D simulations and the 2D simulation, we notice the presence of two distinct instabilities, similar to the observations reported in the benchmark publication [18]. The peak in the FFT plots at $\frac{k_z}{k_0} \sim 0.2$ is associated with the azimuthal component of the MTSI [18], whereas the peak at $\frac{k_z}{k_0} \sim 1$ corresponds to the fundamental ECDI mode [18]. The second peak at the $\frac{k_z}{k_0}$ slightly higher than 2 is related to the second harmonic of the ECDI. Comparing the FFTs from the high-order quasi-2D simulations with respect to those from the 2D simulation, it is observed that the above three peaks are captured in all reduced-order simulations. Moreover, the FFT plots from the 50- and 100-region simulations closely resemble the FFTs from the 2D simulation within each time interval.

Concerning the FFT plots from the low-order quasi-2D and the 1D azimuthal simulations (Figure 8, bottom row), we see that the fundamental ECDI mode is captured in all simulations. However, comparing the azimuthal FFT from the 1D azimuthal simulation against that from the single-region simulation, the magnitude of the ECDI mode and the overall FFT spectrum from the single-region simulation is observed to be more consistent with the high-order quasi-2D and 2D simulations' FFTs. The 3-region and 5-region simulations also resolve the second harmonic of the ECDI. The MTSI is noticed to be absent in the 1D azimuthal and low-order quasi-2D simulations.

The FFT analysis results presented above confirm that the azimuthal wave content from the quasi-2D simulations is quite similar to that from our full-2D simulation. It is discussed in Ref. [18] based on results of full-2D simulations that the existing ECDI and MTSI modes exhibit a nonlinear coupling throughout the simulation, which is the reason behind the oscillatory evolution of the radial electron temperature (also the ion number density albeit to a lesser extent) at the system's quasi-steady state. Accordingly, the discharge in the present radial-azimuthal simulation case is reported to oscillate between two distinct states corresponding to the local maxima and the local minima of the electron temperature evolution plot (Figure 6(f)) [18]. When the radial electron temperature is at its local maximum, the MTSI is dominant, whereas the ECDI becomes more dominant when the radial electron temperature reaches a local minimum.

These insights from conventional full-2D simulations are also recovered using the high-order quasi-2D simulations. Indeed, looking at Figure 9 and Figure 10, which show the 2D snapshots of several plasma properties at a time instance corresponding to a local maximum and a local minimum of the radial electron temperature, respectively, it is evident that with as few number-of-regions as 10, the resolved 2D distributions are in a good agreement compared to the distributions from our full-2D simulation. Moreover, as it might be expected based on the results presented so far, the 50- and 100-region simulations provide predictions of the 2D distributions of the plasma properties which are almost indistinguishable compared to the full-2D results. In this respect, it is particularly seen that both the azimuthal and radial wavenumber components of the MTSI are resolved, which is best visible from the 2D snapshots of the axial electron current density ($J_{ey}$) on the third row of Figure 9.

Of course, moving to the lower end of the high-order reduced-order simulations, i.e., the 10-region and 25-region simulations, some of the small-scale features observable from the other simulations are absent, for instance, the tilts in the plasma distributions near the walls. However, the overall patterns and the main characteristics are still faithfully captured by these relatively low-cost reduced-order kinetic simulations.

To conclude the discussions in this section, we first refer to Figure 11(a), which shows the variation in error of the quasi-2D predictions in terms of the value of the time-averaged ion number density and electron temperature at the mid radial plane with respect to the corresponding full-2D values. It is overall noticed that, from the 3-region quasi-2D simulation, the error in both plasma parameters continuously decreases. In particular, the high-order quasi-2D simulations (25, 50 and 100 regions) have all an error of below 20% in terms of both plasma properties. In addition, it is interesting to note that, although the single-region simulation has quite a large error in terms of the ion number density at the mid radial plane, its predicted radial electron temperature is almost the same as the 100-region simulation. This observation warrants further investigation into the predictions' accuracy of the reduced-order simulations, especially in the low-order approximation limit, in a simulation case with self-consistent ionization. This investigation is left for future work.

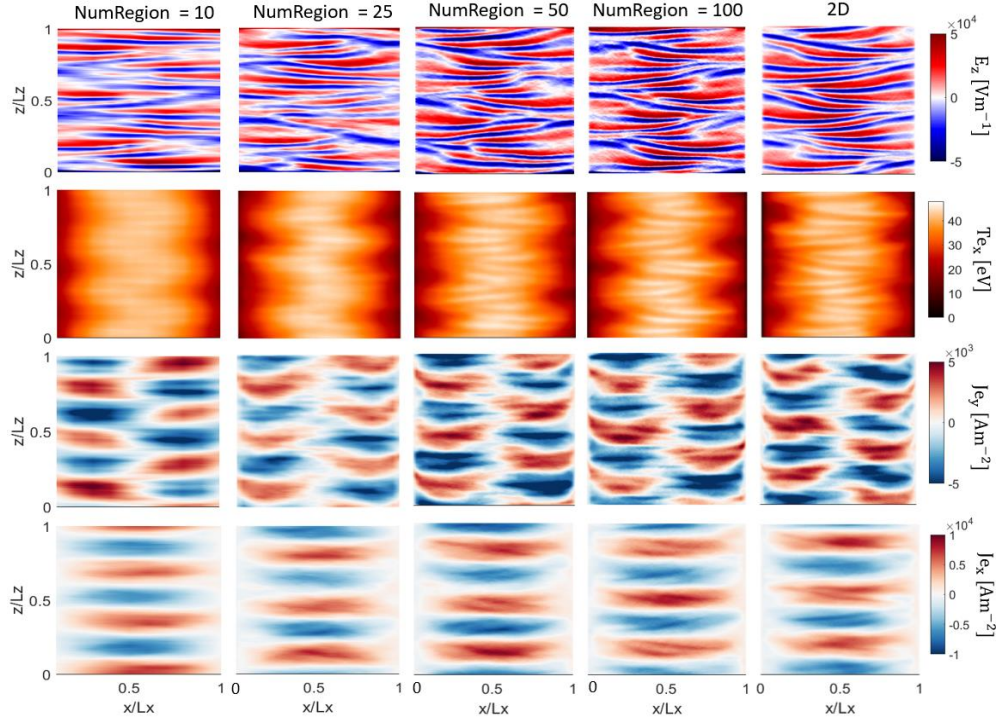

Figure 9: 2D snapshots of the plasma properties at the time of local maximum of the radial electron temperature obtained from high-order quasi-2D simulations and the 2D simulation. The rows, from top to bottom, represent the azimuthal electric field ($E_z$), the radial electron temperature ($T_{ex}$), and the axial and radial electron current densities ($J_{ey}$ and $J_{ex}$).

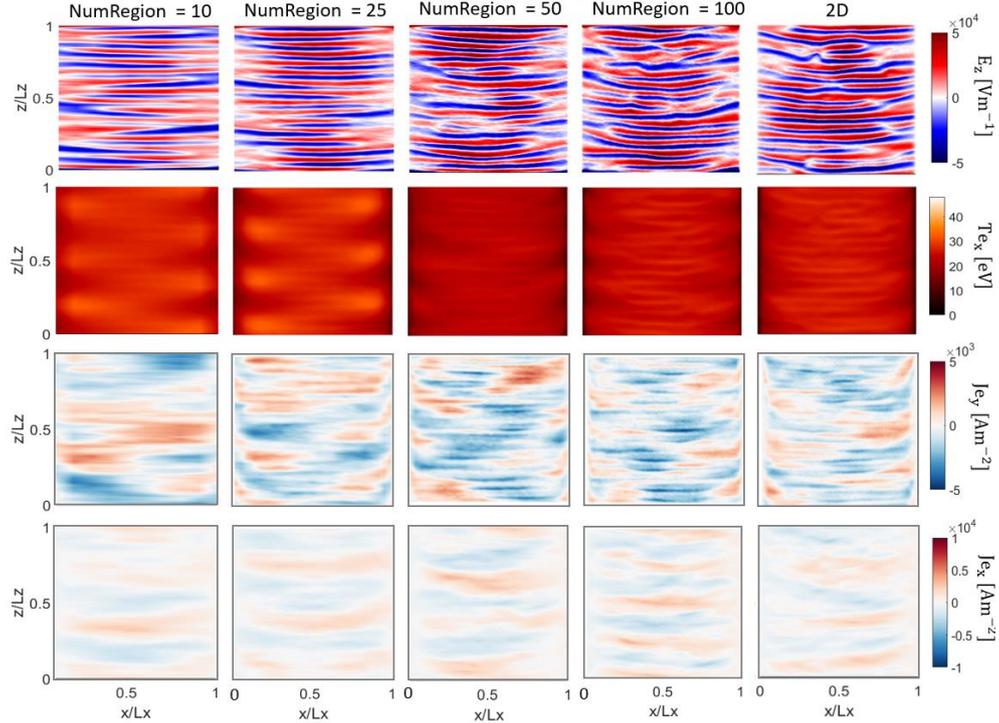

Figure 10: 2D snapshots of the plasma properties at the time of local minimum of the radial electron temperature obtained from high-order quasi-2D simulations and the 2D simulation. The rows, from top to bottom, represent the azimuthal electric field ($E_z$), the radial electron temperature ($T_{ex}$), and the axial and radial electron current densities ($J_{ey}$ and $J_{ex}$).

We now look at Figure 11(b), where we have plotted the ratio between the computational time of the quasi-2D simulations and the 2D one vs the number of regions used for the quasi-2D simulations. This plot, together with the error plot in Figure 11(a), underlines the cost-effectiveness and fidelity of the reduced-order PIC scheme. Indeed, the highest order quasi-2D simulation carried out in this work, i.e., the 100-region simulation, has a computational time that is only 40 % of the 2D one and provides results that are within 2.5 % of the full-2D results. The 50-region simulation with a maximum error of about 10 % in $n_i$ and 5 % in $T_e$ has a computational time ratio of 20 % compared to the 2D simulation. The computational time ratio decreases monotonically to below 5 % for the 10-region simulation for which the error remains below 25 % in terms of both $n_i$ and $T_e$.

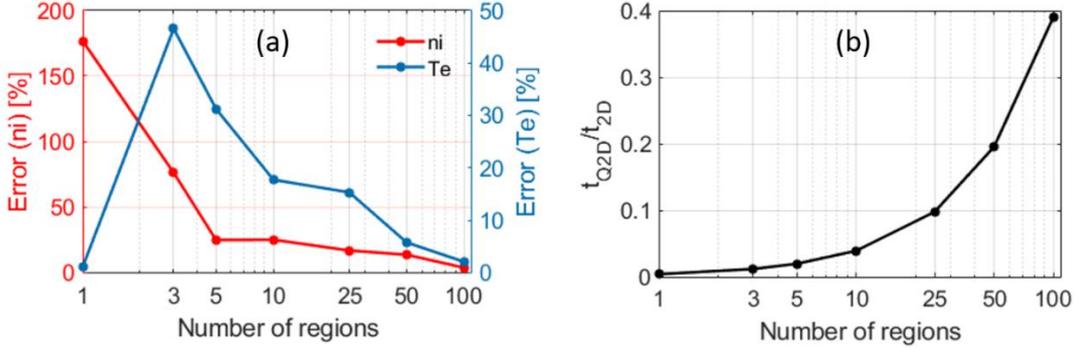

Figure 11: (a) Variation in error of the quasi-2D predictions vs the number of regions used. The errors are represented in terms of the value of the ion number density and electron temperature at the mid radial position compared to the corresponding 2D values; (b) Variation in the computational time ratio between the quasi-2D and full-2D simulations vs the number of regions used in the quasi-2D simulations.

**Section 5: Analysis of the effect of using non-uniform regions' extent in the reduced-order simulations**

The quasi-2D simulations, whose results were presented and discussed so far, all featured a uniform regions' extent along the radial and azimuthal directions. In this section, we investigate the variation in the quasi-2D predictions in case of adopting non-uniform regions' extent along the radial direction. This is to assess that, by using more refined regions near the walls, where the potential gradient is strong due to the plasma sheath, whether a reduced-order simulation with an overall number of regions quite lower than 100 can provide results that resemble those from the 100-region quasi-2D simulation. In the studies presented in this section, the regions along the azimuthal direction are always spaced uniformly.

In this regard, Figure 12(a) shows the schematics of a 2D domain with a non-uniform domain decomposition along the x-direction. Figure 12(b) compares the radial extent of the regions in the four cases investigated here against three reference quasi-2D simulations with uniform regions' definition, namely, the 10-, 25-, and 100- region simulation. Cases 1 and 2 have a total number-of-regions count along the radius (M) of 25, whereas, in Cases 3 and 4, M is equal to 30. It is observed from Figure 12(b) that, in all cases, the smallest radial regions' extent near the walls is equivalent to that of the 100-region simulation. Moving toward the center of the domain along the radial direction, Cases 1 and 2 have a middle region whose radial length is larger than that in the 10- region simulation. In Cases 3 and 4, the middle regions have a radial length larger than that in the 25-region simulation but smaller than the 10-region.

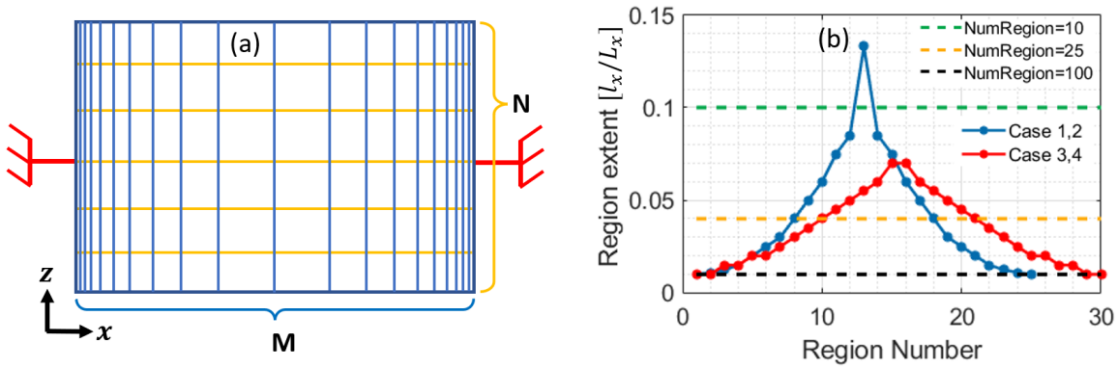

Figure 12: (a) Illustration of non-uniform regions' extent along the radial direction, (b) Comparison of the variation in regions' extent in the different cases investigated.

Table 2 provides a summary of the number of radial (M) and azimuthal (N) regions used in the quasi-2D simulations of this section alongside the initial total number of macroparticles ($N_{p,init}$) and the initial macroparticles-per-cell count ($N_{ppc}$) in each case. It is noticed from Table 2 that Cases 1 to 3 all have 10 regions along the azimuth whereas Case 4 has 100 azimuthal regions. The difference between Case 1 and 2 is the initial total number of macroparticles and, consequently, the $N_{ppc}$. It is recalled that, in the reduced-order PIC scheme, the number of cells to be used for the calculation of $N_{p,init}$ is obtained from the relation $max\,(N*N_i\,,\,M*N_k)$, as described in Ref. [13]. Accordingly, in Case 1, we have exactly 100 macroparticles at the beginning of the simulation in the smallest cells adjacent to the walls. In Case 2, however, there are 100 macroparticles per cell on average, which implies a $N_{ppc}$ of 25 in the smallest cells near the walls. The difference between Cases 3 and 4 is in the number of azimuthal regions (N). The difference in terms of $N_{ppc}$ is a consequence of this variation.

|  | M (nonuniform) | N (uniform) | $N_{p,init}$ | $N_{PPC}$ (In smallest cell) | $N_{PPC}$ (In largest cell) |
|---|---|---|---|---|---|
| **Case 1** | 25 | 10 | $100*max\,(N*N_i\,,\,100*N_k)$ | 100 | 1330 |
| **Case 2** | 25 | 10 | $100*max\,(N*N_i\,,\,M*N_k)$ | 25 | 332 |
| **Case 3** | 30 | 10 | $100*max\,(N*N_i\,,\,M*N_k)$ | 33 | 231 |
| **Case 4** | 30 | 100 | $100*max\,(N*N_i\,,\,M*N_k)$ | 100 | 700 |

Table 2: Summary of the numerical parameters used for the quasi-2D simulations with non-uniform radial regions' extent

The results from the non-uniform quasi-2D simulations in terms of the time-averaged profiles of the ion number density ($n_i$), the total electron temperature ($T_e$), and the radial and azimuthal components of the electron temperature ($T_{ex}$ and $T_{ez}$), given by Eqs. 3 and 4, are compared in Figure 13 against the 10-region, 25-region, and 100-region quasi-2D simulations with uniform radial domain decomposition. It is, overall, observed in the plots of Figure 13 that, by refining the regions' extent near the walls, the ion number density (Figure 13(a)) and the radial electron temperature (Figure 13(c)) distributions closely follow the 100-region profiles shown as dashed black lines in Figure 13. In particular, Case 3 (solid green lines) with only 30 radial regions can reproduce with good accuracy the 100-region results in terms of $n_i$ and $T_{ex}$ profiles.

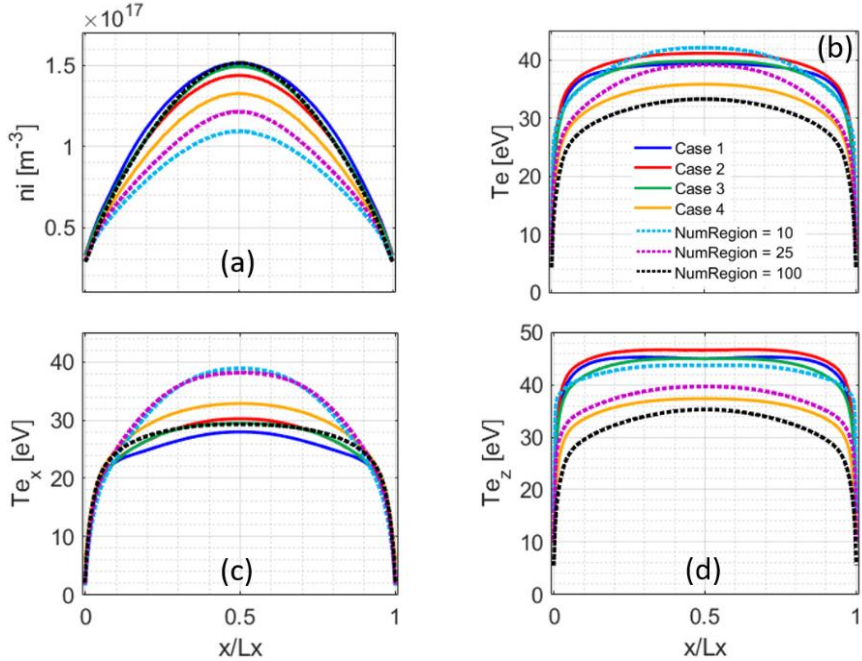

Figure 13: Radial profiles of (a) ion number density, (b) total electron temperature, (c) radial electron temperature, and (d) azimuthal electron temperature from the non-uniform and uniform quasi-2D simulations, averaged over 25-30 $\mu s$.

Nevertheless, the $T_e$ and $T_{ez}$ profiles are seen in plots (b) and (d) of Figure 13 to only approach the 100-region results in Case 4 where the number of azimuthal regions is increased to 100. Indeed, the $T_{ez}$ profiles for Cases 1 to 3 with 10 azimuthal regions are similar to the $T_{ez}$ profile from the 10-region quasi-2D simulation (dashed blue lines) and, hence, the $T_e$ profiles for these cases are also representative of that in the 10-region simulation. Having a fine regions' resolution along the azimuth in Case 4 yields improved predictions of $T_e$ and $T_{ez}$, whereas the $n_i$ and $T_{ex}$ profiles are noticed to deviate to some extent from the 100-region profiles. The reason for this observation concerning Case 4 is thought to be likely due to the influence of the number of macroparticles per cell. We will assess the effect of the $N_{ppc}$ on the quasi-2D simulation results is the following section.

Comparing the profiles of the plasma properties from Case 1 and 2 in Figure 13, it is noticed that the distributions are almost identical, particularly in terms of $n_i$ and $T_{ex}$. Noting that the difference between these two cases was related to the initial number of macroparticles, specifically in the regions near the walls (Table 2), we conclude that a lower $N_{ppc}$ in the smallest regions does not have a noticeable influence on the results.

Finally, looking at Figure 14, which shows the 2D snapshots of the plasma properties for Cases 2 to 4, at a time instance corresponding to the local maximum of the radial electron temperature, we can confirm the conclusions drawn above from the time-averaged profiles. It is obvious from the 2D distributions in Figure 14 that refining the regions near the walls has overall improved the ability of the relatively low number-of-region quasi-2D simulations to resolve the small-scale features and patterns.

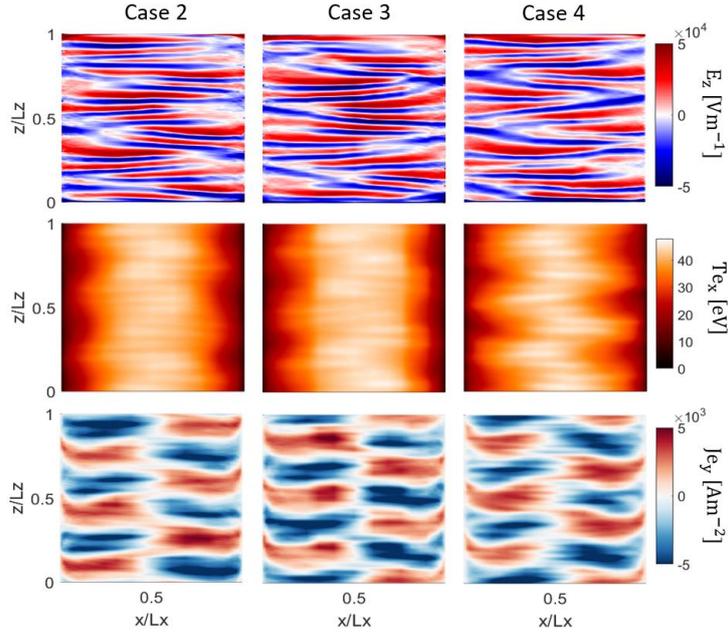

Figure 14: 2D snapshots of the plasma properties at the time of local maximum of the radial electron temperature obtained from the non-uniform quasi-2D simulations (Cases 2 to 4). The rows, from top to bottom, represent the azimuthal electric field ($E_z$), the radial electron temperature ($T_{ex}$), and the axial electron current density ($J_{ey}$).

The analyses presented in this section was primarily meant as a demonstration of the flexibility of the reduced-order PIC scheme to allow for a non-uniform regions' extent, and to specify that, in case the domain decomposition is informed by a knowledge of the macroscopic gradients in plasma properties, particularly the electric field, the overall number of regions can be reduced without a notable impact on the fidelity of the results.

**Section 6: Analysis of the effect of the initial number of macroparticles per cell on the quasi-2D radial-azimuthal simulation**

To evaluate the influence of the initial number of macroparticles per cell on the quasi-2D results, we used the uniform 50-region quasi-2D simulation, which was shown in Section 4 to provide high-fidelity results at relatively low computational cost, to simulate the radial-azimuthal case with a range of initial $N_{PPC}$, from 10 to 1024. The results in terms of the normalized, time-averaged, radial distributions of the ion number density and total electron temperature are shown in Figure 15. Referring to plot (a) in this figure, the ion number density profiles are seen to have converged at a peak density ratio of about 2.45 from 100 macroparticles per cell. Looking at plot (b) of Figure 15, even though the convergence trend is not monotonic, the temperature profiles does not show any noticeable variation from an initial $N_{PPC}$ of 200, and the simulation case with 100 $N_{PPC}$ has a maximum discrepancy of less than 10 % in normalized $T_e$ compared to the simulations with higher $N_{PPC}$ counts.

As a result, it is possible to say that the quasi-2D simulations show a clear statistical convergence behavior when increasing the initial $N_{PPC}$, and an initial macroparticle per cell in the range of 100 to 200 can be sufficient to ensure statistically converged results. This conclusion is consistent with that from the similar studies reported in the literature [18][28] and also from our previous works [13][14].

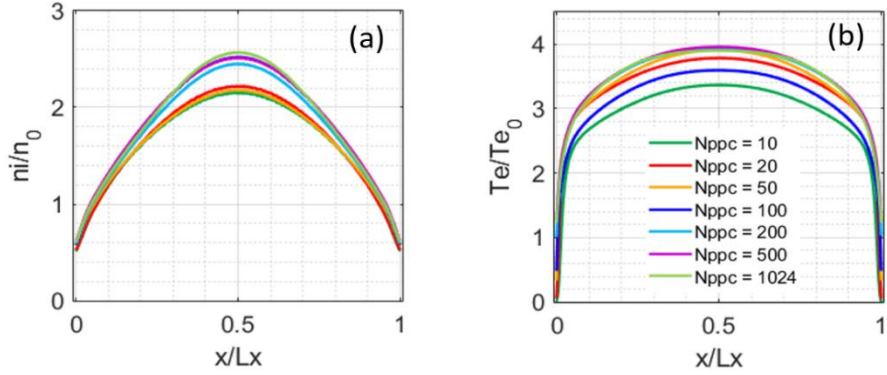

Figure 15: Time-averaged radial profiles of (a) normalized ion number density, and (b) normalized electron temperature obtained from the uniform 50-region quasi-2D simulation with various initial number of macroparticles per cell. $n_0$ and $T_{e0}$ denote the initial ion number density and electron temperature used for initial particles' loading.

**Section 7: Conclusions**

In this article, we presented the verification results of two PIC codes, the conventional IPPL-2D and the reduced-order IPPL-Q2D, in a radial-azimuthal E × B configuration representative of a Hall thruster. The simulation setup and conditions were based on a well-defined benchmark case in the same configuration [18]. We introduced the IPPL-2D code and presented its benchmarking against the 2D results from the literature. Accordingly, we confirmed that the full-2D results can be used as a trustworthy reference for the verification of the reduced-order PIC code. In this respect, we first provided an overview of the reduced-order PIC scheme and highlighted the two aspects with respect to which the resulting IPPL-Q2D code is different from IPPL-2D, namely, the formulation behind the Poisson solver, and the computation grids used for the information exchange between the particles and the grids.

Second, we presented an extensive comparison between the quasi-2D simulation results at various approximation orders (or, equivalently, different number of regions) and the full-2D results. We compared the quasi-2D and full-2D predictions in terms of the time-averaged radial profiles of the plasma properties, the time evolution of the plasma properties, 1D FFTs of the azimuthal electric field signals, and, finally, the 2D snapshots of the plasma properties at two instances of the discharge evolution at the quasi-steady state corresponding to a local minimum and a local maximum of the radial electron temperature. Considering the highly two-dimensional nature of the plasma phenomena in the radial-azimuthal coordinates of a Hall thruster, we observed that reduced-order simulations at the approximation orders that yield from about 2.5 to 25 times speed-up compared to the 2D simulation provide results that compare reasonably well with the full-2D ones. In particular, we demonstrated that a quasi-2D simulation with 50 regions along the radial and azimuthal directions has a speed-up factor of 5 while its predictions in terms of the time-averaged plasma profiles and the 2D distributions of the plasma properties are well consistent with the corresponding predictions from the 2D simulation.

On the highly low-cost end of the reduced-order approximations, the single-region simulation, which practically amounts to in-parallel 1D PIC simulations along each of the simulation coordinates, was shown to provide results that are notably improved with respect to the corresponding 1D radial and 1D azimuthal simulations. This improvement in predictions was underlined to be due to an average mutual effect of the radial and azimuthal physical processes that is resolved by the single-region simulation.

We also investigated the effect of using non-uniform domain decomposition along the radius in the quasi-2D simulations as well as the influence of the initial number of macroparticles per cell. Concerning the former, we confirmed the flexibility of the reduced-order formulation to allow for arbitrary selection of the regions' extent. In addition, we highlighted that refining the domain decomposition around the locations of large gradient in the plasma properties can enable a non-uniform quasi-2D simulation with relatively low number of regions to have a results' fidelity on a par with the higher-order quasi-2D simulations with uniform, refined regions' extent throughout the domain.

The results and in-depth analyses presented in this paper complement our previous detailed evaluations of the reduced-order scheme in the axial-azimuthal coordinates of a Hall thruster [12][13]. As a result, the generalized reduced-order PIC scheme is believed to be mature, versatile, and reliable enough so that, in the near term, it can be used for extensive physical parametric studies that are essential to better understand and untangle the roles played by various mechanisms in the overall picture of the multi-dimensional processes in Hall thrusters and similar cross-field plasma configurations. The results from a series of such studies that we have carried out in a radial-azimuthal configuration are currently being finalized and will be reported in a follow-up publication.

**Acknowledgments:**

The present research is carried out within the framework of the project "Advanced Space Propulsion for Innovative Realization of space Exploration (ASPIRE)". ASPIRE has received funding from the European Union's Horizon 2020 Research and Innovation Programme under the Grant Agreement No. 101004366. The views expressed herein can in no way be taken as to reflect an official opinion of the Commission of the European Union.

**Conflict of Interest:**

The authors have no competing interests to declare that are relevant to the content of this article.

**Data Availability Statement:**

The data that support the findings of this study are available from the corresponding author upon reasonable request.